# Identification of potential Music Information Retrieval technologies for computer-aided jingju singing training

Rong Gong, Xavier Serra
Music Technology Group, Universitat Pompeu Fabra, Barcelona

## Abstract

Music Information Retrieval (MIR) technologies have been proven useful in assisting western classical singing training. Jingju (also known as Beijing or Peking opera) singing is different from western singing in terms of most of the perceptual dimensions, and the trainees are taught by using mouth/heart method. In this paper, we first present the training method used in the professional jingju training classroom scenario and show the potential benefits of introducing the MIR technologies into the training process. The main part of this paper dedicates to identify the potential MIR technologies for jingju singing training. To this intent, we answer the question: how the jingju singing tutors and trainees value the importance of each jingju musical dimension—intonation, rhythm, loudness, tone quality and pronunciation? This is done by (i) classifying the classroom singing practices, tutor's verbal feedbacks into these 5 dimensions, (ii) surveying the trainees. Then, with the help of the music signal analysis, a finer inspection on the classroom practice recording examples reveals the detailed elements in the training process. Finally, based on the above analysis, several potential MIR technologies are identified and would be useful for the jingju singing training.
**Keywords:** jingju, computer-aided singing training, music information retrieval

## 1. Introduction

Jingju singing is traditionally taught between tutor and trainee by using the mouth/heart (oral teaching, 口传心授) and face-to-face methods—"Jingju tuition requires demonstration, and tutors tell trainees the secrets for certain skills that they learned from their own masters or that they worked out from their experience. The close relationship of the tutor-trainee or the master-disciple is based on the mouth/heart teaching method that stresses through oral instruction and intuitive understanding. Imitation is certainly the first step, and it is crucial for our learning process… not even one component in the 'four skills (singing, speech, dance-acting, combat)' can be learned by trainee himself. Much of the nuance of the singing can only be learned from face-to-face teaching." [1]

After five months research stay in NACTA (National Academy of Chinese Theatre Arts, leading institute in China dedicated to the training of professionals in performing and studying traditional Chinese opera), we had a firsthand experience of the month/heart teaching method of jingju singing. In class, the tutor teaches several melodic lines selected from an aria. In the beginning, tutor firstly makes a short introduction about the teaching content - the aria story, the characters, the special pronunciations, etc. Then she/he gives a demonstrative singing of these lines. In the rest part of the class, the trainees are asked to imitate the demonstrative singing line by line, and the tutor corrects the imitations. The process of this part can be generalized as: (i) the tutor asks the trainees to give a tentative singing individually or in chorus for either a line, or several syllables, or only one syllable. (ii) Then the tutor identifies the singing problems, (iii) gives verbal feedback or a demonstrative singing feedback. (iv) Finally, the trainees does a revised singing with the feedbacks in mind. The step from (ii) to (iv) could be iterated until the trainee's singing satisfies the tutor's criteria. We name one single such process as **a correction occurrence** (see figure 1).

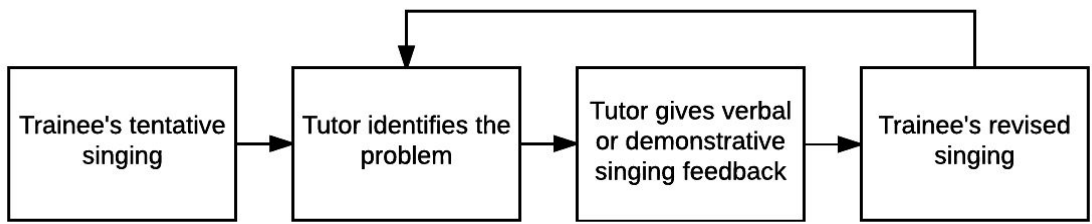

Figure 1. The flowchart of a single correction occurrence.

The verbal feedback is a semantic comment given by the tutor. It's the description of helping the trainee to improve her/his singing performance and the most important information which can clarify the singing problems.

During the research stay, we found that trainees are not able to, more than few times, get the point of tutor's verbal feedback or demonstrative singing. One possible reason could be that tutor's verbal feedback is too personal to be understood by the trainee. One such example is "don't be sloppy, sing it with solidity." The abstract word in this feedback—"sloppy" is easy to confuse the trainee if she/he doesn't have enough training experience with the tutor or is not able to relate these words to some specific singing methods. Another possible reason could be that the trainee's level is not enough for her/him to understand the jingju jargons or perceive the nuance in the demonstrative singing. Callaghan J. etc. mentioned this problem in his paper [2]—"between teacher and student there may well be disparities in perceptual style, and teaching/learning style, in addition to great differences in life experience, making allusive language at best ambiguous, and at worst, misleading." Consequently, it's necessary to bring additional method to assist the month/heart teaching.

The Music Information Retrieval (MIR) technologies have the potential ability to represent music concepts as data or graph for trainee's tentative singing or tutor's demonstrative singing. The data and visual representations serve as complementary information in addition to tutor's verbal feedback, and instruction tools which can be used to explain the feedback concretely—visual feedback, such as pitch contour, has been proved effective in assisting the singing training [3]. By combining with the visual feedback, trainee and tutor are likely to identify the singing problems easier and faster.

Callaghan, J. et al. [4] borrowed some real-time speech visualization technologies, such as pitch trace, spectrogram, vowel chart, and then evaluated them by singing tutors and trainees in singing training situations. Welch, G. F et al.. [5] introduced also several related technologies for the computer-aid singing teaching, such as pitch, spectra, vocal tract area. To evaluate the usefulness of these technologies, they also held a one-day workshop which involves the researchers from speech, singing, psychology, linguistics, vocal health, engineering, and education fields.

The technologies in the previous works are proposed by speech or music technologies engineers, and evaluated by western classical or pop music tutors or trainees. In this study, effort is put on address the problem of identifying potential MIR technologies for computer-aided jingju singing training. In the next chapter, we explain our methods—classifying, finer inspection of tutors' correction occurrence and surveying the trainees. These methods aim to answer the question: how tutors and trainees value the importance of each jingju singing dimensions—intonation, rhythm, loudness, tone quality and pronunciation, and to reveal the detailed jingju singing elements on which tutors lay stress and trainees tend to have problems. In section 3, we report and discuss the analysis results and propose several potential MIR technologies for jingju singing training. We conclude our work in section 4.

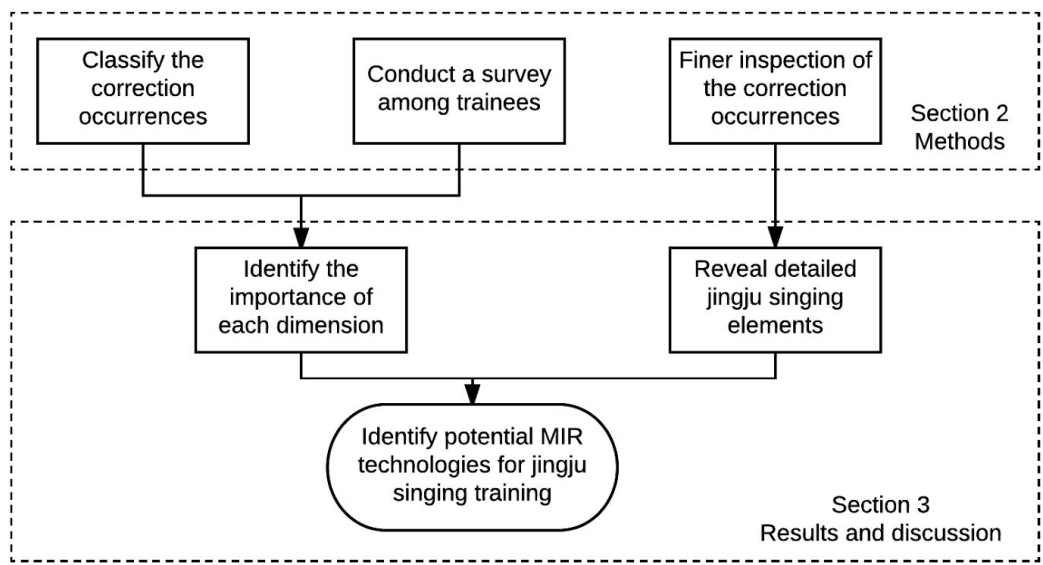

Figure 2. The flowchart of the paper structure.

## 2. Methods

In paper [6], the musicians have rated the performance of western arias in 5 musical dimensions: phrasing/expression, intonation, rhythm, loudness, tone quality. In our study, we borrow the concept of musical dimensions for music performance assessment and adapt them according to jingju singing background.

Almost all jingju aria contains lyrics, and as we will prove in later chapters—to be able to pronounce the singing lyrics accurately is a key skill in jingju singing, we thus add the pronunciation as an independent dimension to above dimension set. Besides, we discard phrasing/expression because it is a "meta-dimension" constructed above other basic dimensions—"A musician accomplishes this by interpreting the music, from memory or sheet music, by altering tone, tempo, dynamics, articulation, inflection, and other characteristics"[1]. Overall, 5 dimensions will be taken into account in this paper—intonation, rhythm, loudness, tone quality and pronunciation. Accordingly, we give their definitions:

- Intonation: accuracy of pitch in singing.
- Rhythm: singing a rhythmic pattern on time. This means that the notes or syllable are not ahead of the time or behind the time.
- Loudness: the loudness dynamic variation between notes/syllables or phrases.
- Tone quality: the color or timbre of the singing voice.
- Pronunciation: the act or result of producing the sounds of speech, including articulation and stress.

By classifying the correction occurrences, we will find out the dimensions on which tutors lay stress or trainees tend to have problems. On the other hand, we conduct a simple survey to investigate the importance of each dimension from trainees' perspective.

Although a general view of jingju singing training can be built upon the 5 dimensions mentioned above, it's still too vague to identify the specific MIR technologies which can help the training process. Some further questions still need to be answered. For example, What is the primary jingju singing time unit—syllable or note? Is the pitch ornament recognition and analysis is useful for revealing the singing problems in the intonation dimension? To answer these questions, we conduct a finer inspection on each correction occurrence.

## 2.1 Correction occurrence analysis

During the research stay in NACTA, we audited and recorded the audio from three singing classes. Three class was taught respectively by three professional tutors, which contain solo and chorus practices. We recorded the class teaching and practicing audio content by using a SONY PCM-D50 stereo portable audio recorder. Only the solo practices are kept for the analysis because they can reveal the singing problems of an individual trainee, whereas the individual voices are blurred in the chorus practice recordings. The audio excerpts of each correction occurrence are then edited and visualized by signal processing tools—pitch contour, loudness contour and spectrogram, which is helpful in identifying the singing problem, especially when the tutor's verbal feedback is too abstract to extract any effective information.

### 2.1.1 Collecting correction occurrence

---

[1] https://en.wikiquote.org/wiki/Musical_phrasing Retrieved 30 September 2017

The table 1 depicts the information of aria name, role-type (four major roles-Sheng (生), Dan (旦), Jing (净), Chou (丑) existing in jingju), trainee number in the class, melodic line number practiced in the class and correction occurrence number collected from the recordings. An example of reading this table is: Three trainees were involved in the laosheng class WuJiaPo. 11 melodic lines were taught and 20 correction occurrences were collected from the recordings.

Table 1. The statistics of the correction occurrence analysis materials

| Aria name | Role-type | Trainee number | Melodic line number | Correction occurrence number |
|---|---|---|---|---|
| 武家坡 (WuJiaPo) | 老生(laosheng) | 3 | 11 | 20 |
| 太真外传 (TaiZhenWaiZhuan) | 青衣(qingyi) | 3 | 3 | 21 |
| 捉放曹 (ZhuoFangCao) | 花脸(hualian) | 2 | 28 | 21 |

The ratios between melodic line number and the correction occurrence number are widely different for the three classes (Table 1). For example, during the TaiZhenWaiZhuan class, 3 trainees practiced 3 lines and were corrected 21 times, which results in a ratio of 1/7. However, during the ZhuoFangCao class, 2 trainees practiced 28 lines and also were corrected 21 times, which has a ratio of 4/3. The correction frequency depends on several factors, such as the trainees' singing levels, the tutor's teaching method. The low singing level trainees tend to receive more corrections than those who have high singing levels.

### 2.1.2 Analysis examples

For each occurrence, we analyze the target recordings and tutor's verbal feedback. Additionally, to achieve the visual analysis, their pitch, loudness contours and spectrogram are also presented.

We firstly classify the correction occurrences into 5 dimensions—intonation, rhythm, loudness, tone quality and pronunciation. A correction occurrence can be classified into more than one dimension. For example, the correction with the verbal feedback "don't be sloppy, sing it with solidity, make the tone quality sounds round." can be classified into intonation (irregular vibrato), loudness (unstable loudness contour) and tone quality (higher harmonics too clear), by analysing comparatively between the tutor's demonstrative singing and trainee's tentative singing. Furthermore, a finer inspection of each correction occurrence is conducted, where we identify the detailed elements.

5 correction occurrences are taken as examples to showcase our analysis. For each one, we list its aria name, melodic line, target syllable, the tutor's verbal feedback, the dimensions classified and the detailed elements identified. Finally, we give a short explanation accompanied with the visualization to justify our classification and identification process.

Occurrence 1:

Aria: TaiZhenWaiZhuan (太真外传)
Line: yi yuan na chai yu he qing yuan yong ding (一愿那钗与盒情缘永定)
Target syllable: yuan (缘)
Tutor's verbal feedback: it didn't jump up. (没跳起来)
Dimension: intonation
Detailed elements: pitch ornament, gliding
Explanation:
The syllable's second tone in the tutor's demonstrative singing has a pitch gliding (ornament). However, the gliding in the trainee's version is not apparent (Fig. 3).

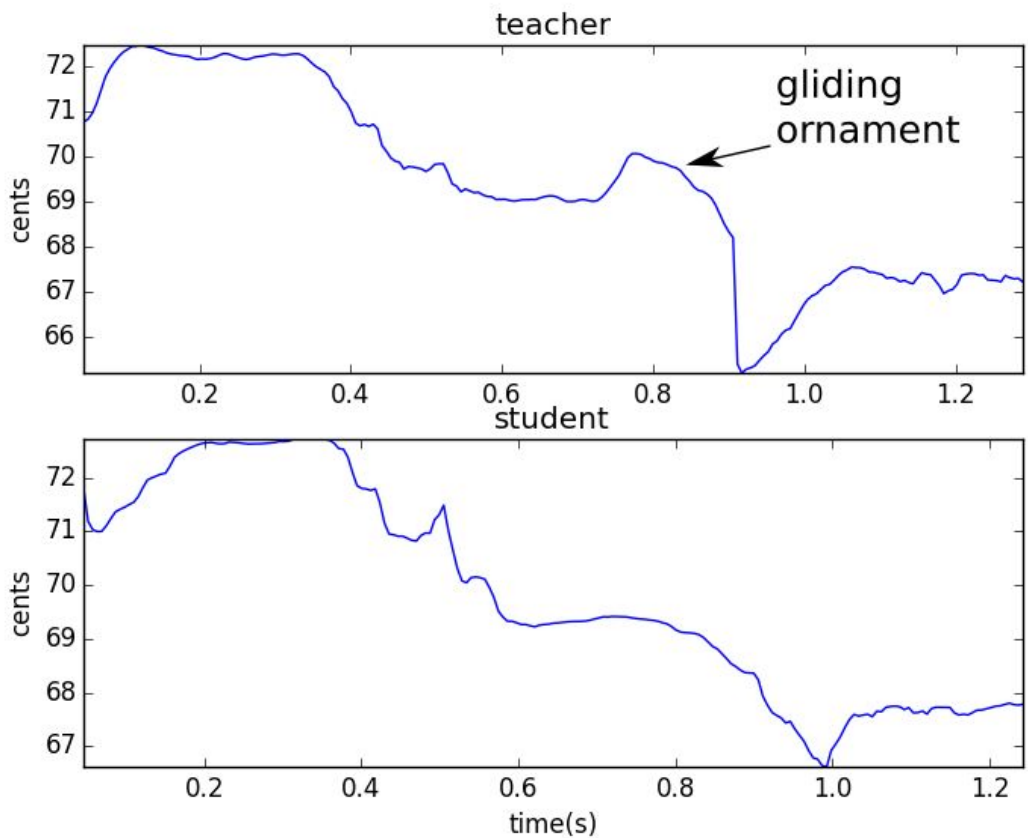

Figure 3. The pitch contours of the syllable "yuan" for occurrence 1.

Occurrence 2:

Aria: ZhuoFangCao (捉放曹)
Line: dang nian jie bai yi lu xiang (当年结拜一炉香)
Target syllables: dang nian jie bai (当点结拜)

Tutor's verbal feedback: swing apart these four syllables (1,2,3,4 四个字甩开)
Dimension: rhythm
Detailed elements: temporal duration of the syllables
Explanation:
In tutor's demonstrative singing, the temporal duration of the third syllable "jie" has been prolonged, in contrast with other three syllables. This can be observed by the pitch contour (Fig. 4).

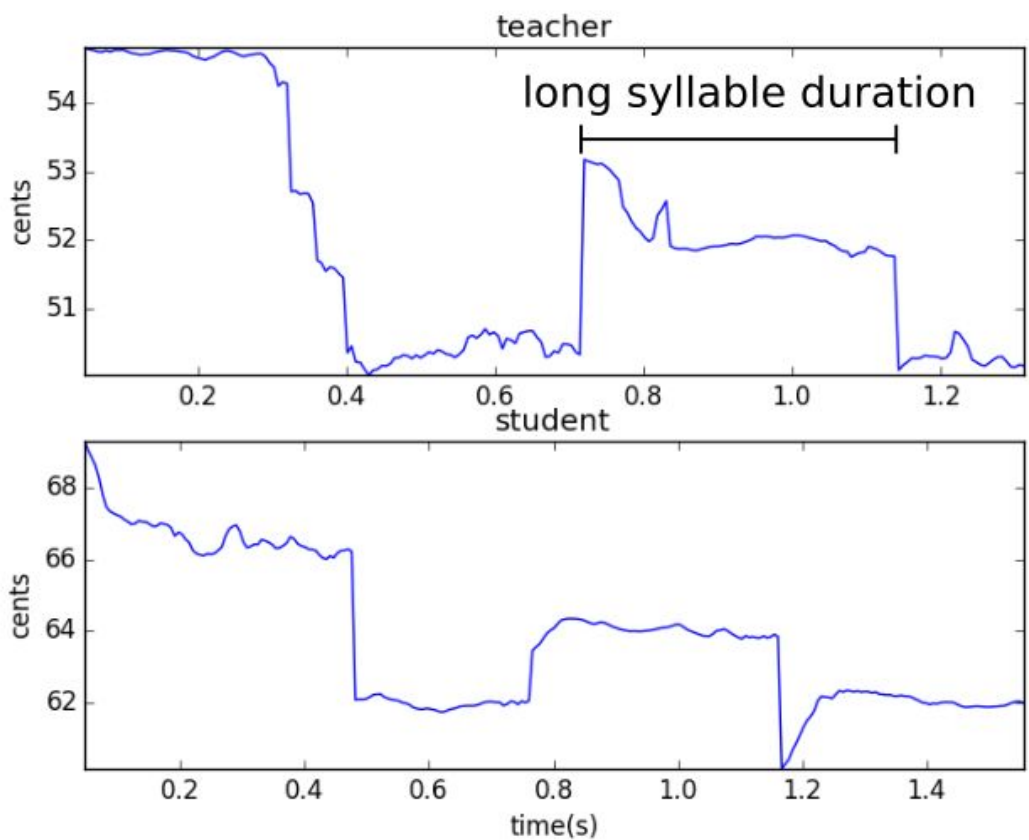

Figure 4. The pitch contours of the syllables "dang nian jie bai" for occurrence 2.

Occurrence 3:

Aria: TaiZhenWaiZhuan (太真外传)
Line: yang yu huan zai dian qian shen shen bai ding (杨玉环在殿前深深拜定)
Target syllable: yang (杨)
Tutor's verbal feedback: emphasizing the nasal voice (an 鼻音要突出)
Dimensions: loudness and tone quality
Detailed elements: specific part of the syllable, harmonics
Explanation:
In tutor's demonstrative singing, a prominent loudness peak can be found in the head of the syllable, which maintains a high loudness level in the belly (Fig. 5). We also can observe that the higher harmonics are abundant from the spectrogram (Fig. 6).

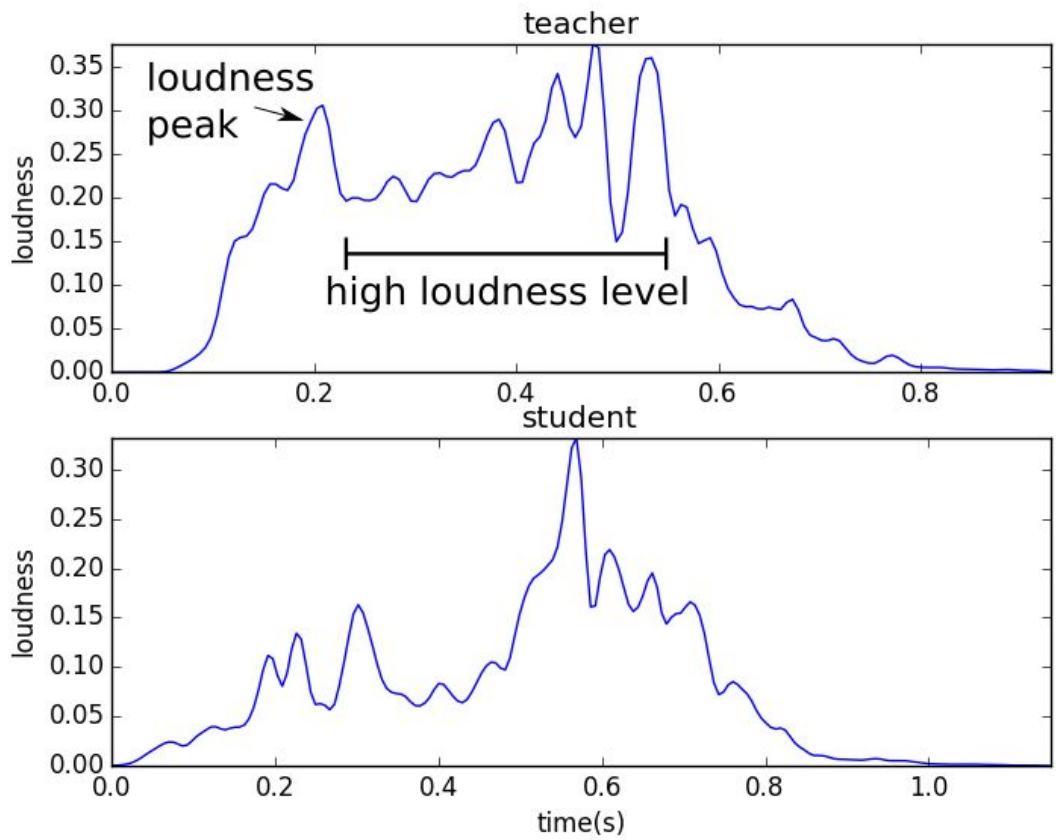

Figure 5. The loudness contours of the syllable "yang" for occurrence 3..

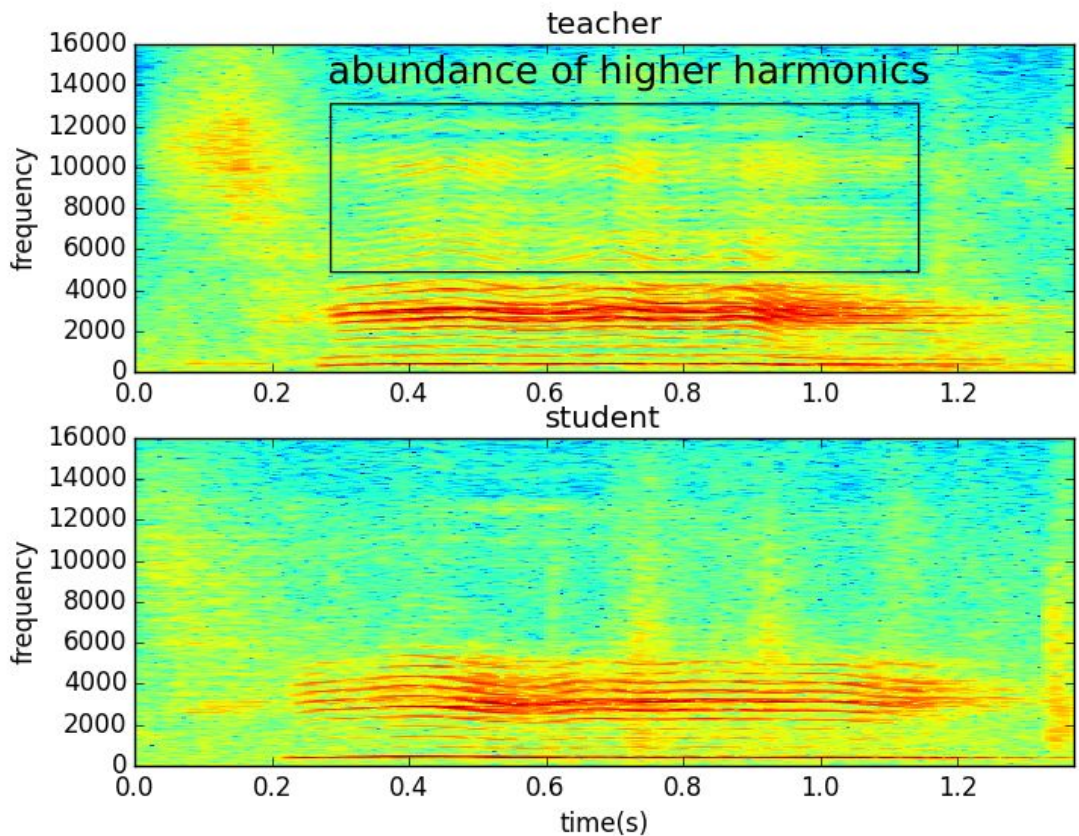

Figure 6. The spectrograms of the syllable "yang" for occurrence 3.

Occurrence 4:

Aria: ZhuoFangCao (捉放曹)
Line: xian xie zuo le na wa shang shuang (险些做了那瓦上霜)
Target syllable: shang (上)
Tutor's verbal feedback: terminate the sound at 'ng' (sound收音收到 ng)
Dimension: pronunciation
Detailed elements: specific part of the syllable, harmonics
Explanation:
The tutor's demonstrative singing is one octave lower than the trainee's singing. The tutor's feedback emphasizes on the goodness of the syllable tail—ng. His demonstrative singing contains fewer harmonics in the tail than the trainee's singing (Fig. 7).

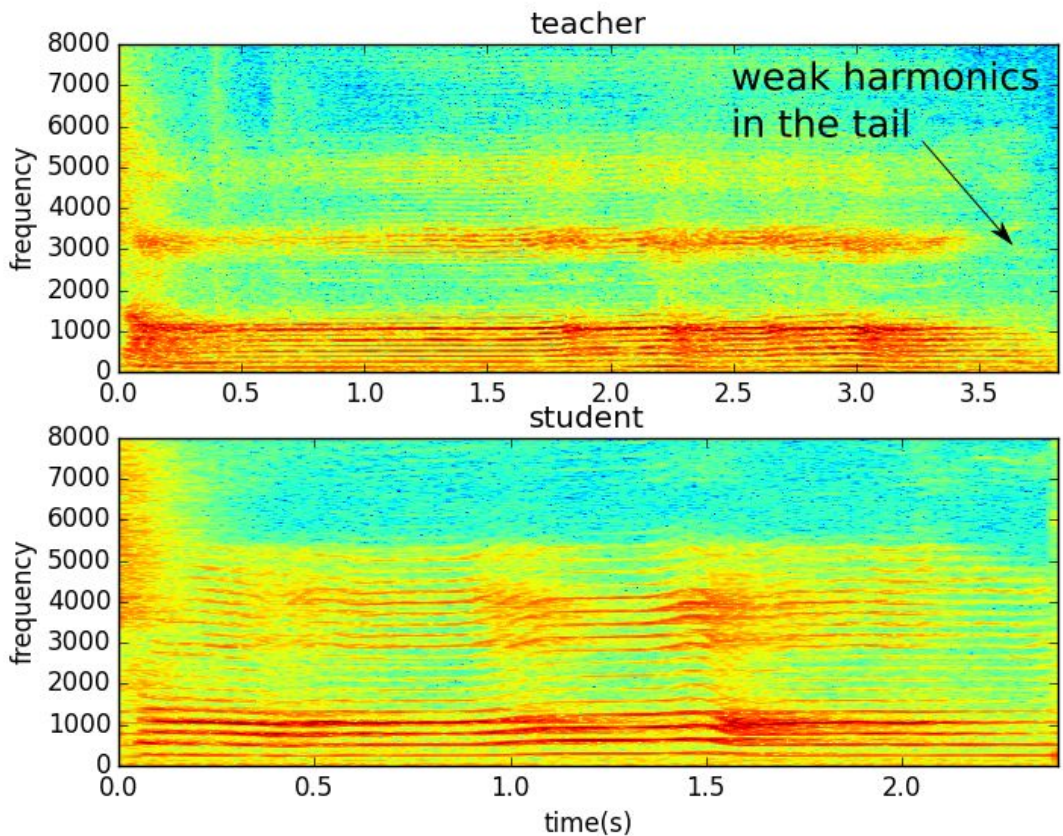

Figure 7. The spectrograms of the syllable "shang" for occurrence 4.

Occurrence 5:

Aria: WuJiaPo (武家坡)
Line: jian liao na zhong da sao xi wen kai huai (见了那众大嫂细问开怀)
Target syllable: kai huai (开怀)
Tutor's verbal feedback: adjust the breath, make the sound solid even if you sing in the low register (要用气，低调们也要放实在了)
Dimension: tone quality
Detailed elements: harmonics, resonance
Explanation:
This feedback has twofold of meaning. First is to take an enough breath, and have an enough air in the chest to sing. Second is to adjust the body's resonance position to make the sound more solid. This can be observed as in the spectrogram of the tutor's demonstrative singing, which contains less energy for the lower harmonics and the prominent resonances (energy) in around 750 Hz and 1250 Hz (Fig. 8).

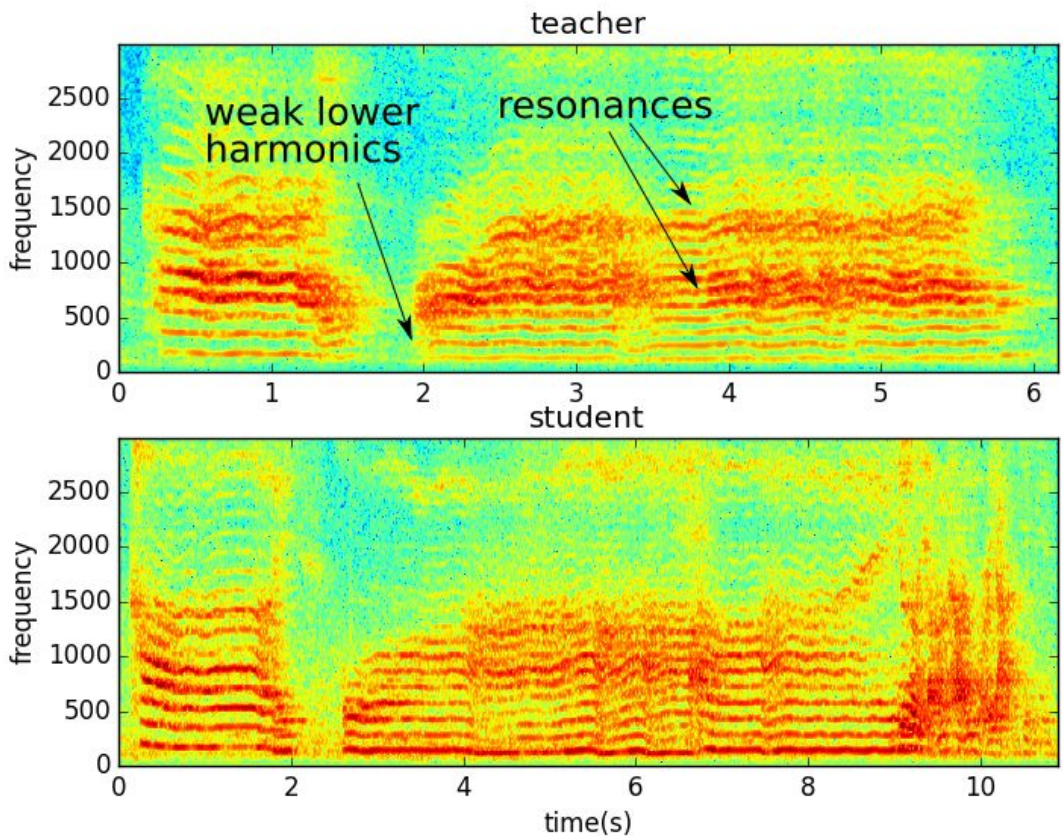

Figure 8. The spectrograms of the syllables "kai huai" for occurrence 5.

## 2.2 A survey among trainees

We conduct a simple survey among another 9 trainees to investigate the importance of each dimension from their perspectives. The survey contains two questions:

1. Please rate the use frequency of the following media when you learn to sing arias - score, audio recording or tutor's classroom teaching.
2. Please rate the importance of the following jingju singing dimensions and elements when you learn to sing arias—intonation, rhythm, loudness, pronunciation, ornament, breath and strength (劲头).

9 trainees have participated in this survey; they are different from the ones presented in section 2. Among them, 5 are professional undergraduate or graduated trainees from NACTA, 4 are amateurs from the jingju associations in two non-art universities in Beijing. We use a five-level Likert scale[2] for each rating term. For example, the "use frequency of the score" in the first question and the "importance of intonation" in the second question can be rated from 1 to 5, where 1 means "never used" or "not important at all" and 5 means "most frequently used" or the "most important". Then, we take the average value of each term for 5 professional trainees and 4 amateurs respectively.

---

[2] https://en.wikipedia.org/wiki/Likert_scale retrieved 30 September 2017

It is worth to mention that three more elements—ornament, breath and strength have been added to the survey. The consideration for this change is that the survey terms need to be adapted to the trainee's artistic background, and the jingju singing jargons should be easily accessible by them.

# 3. Results and discussion

In this section, we report the results of the analysis on tutors' correction occurrences and the trainees' survey. The correction occurrences are classified into 5 dimensions and the detailed elements about each correction occurrence are also analyzed by using the method introduced in the section 2.2.2. Then, we describe the trainee' survey result and compare it with the tutors' correction result. Lastly, we discuss the potential MIR technologies for computer-aided jingju singing training.

## 3.1 Correction occurrence analysis

Table 2. The statistics of the correction occurrence dimension classification.

| Class | Intonation | Rhythm | Loudness | Pronunciation | Tone quality |
|---|---|---|---|---|---|
| 武家坡 (WuJiaPo) | 8 | 0 | 1 | 6 | 6 |
| 太真外传 (TaiZhenWaiZhuan) | 6 | 1 | 9 | 4 | 11 |
| 捉放曹 (ZhuoFangCao) | 5 | 2 | 7 | 9 | 3 |
| Sum | 19 | 3 | 17 | 19 | 20 |

We observe from the Table 2 that among the 5 dimensions, tone quality, pronunciation and intonation dimensions have the largest and almost equal occurrence number, loudness takes the second place, and rhythm problem was least mentioned. In other words, tone quality, pronunciation and intonation are the dimensions which receive particular attention by tutors and cause problems easily to trainees.

By using the method mentioned in section 2.2.2, we identified the following 8 detailed elements which are emphasized by the tutors. These elements are listed by their occurrence number, in the descending order:
1. The specific part of the syllable: problems related to the specific part of the syllable—head, belly and tail (occurrences 3 and 4 in section 2.2.2).
2. Accuracy: problems associated to the singing accuracy, such as intonation, loudness and pronunciation accuracies.

3. Body resonance position: problems associated to the use of the correct resonance position/cavity (occurrence 5 in section 2.2.2).
4. Other tone quality problems.
5. Ornament: singing accuracy issues related to pitch ornaments such as vibrato, gliding (occurrence 1 in section 2.2.2) and pronunciation ornaments such as sou sound (嗽音), houtou sound (喉头音).
6. The specific note of the syllable: problems associated to the specific note within a syllable.
7. Temporal duration of the syllables: problems associated to the syllable durations in a melodic line (occurrence 2 in section 2.2.2).
8. The transition between two parts of the syllable: problems associated to the transition between two parts of the syllable - from head to belly or from belly to tail.

The analysis results are organized to an Excel spreadsheet[3], which consist of the tutor's verbal feedback, signal analysis method, classified dimension and detailed element.

## 3.2 The survey among trainees

We gather the survey results by ordering the mean values for each question. For the first question, the use frequency order from high to low of three learning media are:
(i) Professional group: classroom teaching, audio recording, score;
(ii) Amateur group: audio recording, teacher's classroom teaching, score.

The music score has been rated as the lowest use frequency by both professional and amateur groups, which means that the jingju trainees we investigated do not use the visual clue - music score reading, to learn arias. The tutor's classroom teaching has been rated as the highest use frequency for the professional and the second for the amateurs, which is reasonable because this learning medium is much easier available for the professional. Lastly, the high rating of both tutor's classroom teaching and audio recording shows that the jingju trainees use mostly the listening and imitation methods to learn arias.

For the second question, the importance order from the most important to most trivial are:
(i) Professional group: rhythm, strength, pronunciation, breath, intonation, loudness, ornament;
(ii) Amateur group: rhythm, pronunciation, strength, ornament, breath, intonation, loudness.

Apart from the terms strength and breath, the others have been analyzed in sections 3.1 from the correction occurrence perspective. Strength is a stylistic and abstract word to depict the energy used in jingju singing and instrument playing. A jingju singing with strength is conveyed by combining multiple elements, such as loudness (mostly), rhythm, intonation and tone quality. Breath or specific methods of breathing (气口) described in Wichman [8] is "these methods allow the exiting breath to control the pitch, timbre or tone color, and energy of the sound produced." In consequence, Strength and breath both are nonspecific terms combining or affecting multiple basic jingju singing elements.

---
[3] https://doi.org/10.5281/zenodo.1014025

Pronunciation is rated as an important element by both the professional and amateurs, which is coherent with the result in section 3.1. The high importance of rhythm and low importance of intonation and loudness contradict to the result in section 3.1. For rhythm aspect, one possible explanation is that the higher importance the trainees value a term, less prone they are to sing poorly on it—the trainees considered rhythm is the most important singing aspect, so they are less prone to have the rhythmic problems. For intonation and loudness, we can't easily conclude that they are not important in the learning process. The reasons are twofold: on the one hand, the trainees might think that the intonation accuracy is a basic requirement in jingju singing and its importance is self-evident; on the other hand, because intonation and loudness are jargons used in acoustic, sound and music technology research fields, which might be foreign to these trainees, so they might avoid them and choose the familiar terms such as strength.

## 3.3 Potential MIR technologies

In this section, we summarize the discussion in sections 3.1 and 3.2 and identify the potential MIR technologies for jingju singing training.

The only jingju singing dimension emphasized in both correction occurrence analysis and the survey analysis is pronunciation, which shows that its predominant role in jingju singing training. However, due to the importance contradiction of several other dimensions, such as rhythm, intonation and loudness, revealed in the occurrence analysis and the survey, we attribute equal importance to the other 4 dimensions.

Four detailed elements 1, 6, 7 and 8 discussed in section 3.1 are related to the singing syllable. Therefore, a method of **automatic segmentation of jingju singing into syllables**, which is able to indicate the syllable onset/offset boundaries, will be useful. Two related works [9,10] have been done to explore this technology. The proposed method in [10] estimates the most likely sequence of syllable boundaries given the estimated syllable onset detection function (ODF) and its score.

The assessment can be more precise in syllable part, note and transition time levels, such as the detailed elements 1, 6 and 8. This requires us to develop technologies which can **automatically segment the syllable into parts, parts transitions and notes**.

The detailed element 5 is related to singing ornament. Thus, an automatic jingju ornaments recognition technology will be helpful in disclosing these singing problems. In the paper [11], we propose a technique for this purpose. Pitch contour of each musical note is segmented automatically by a melodic transcription algorithm incorporated with a genre-specific musicological model of jingju singing: bigram note transition probabilities defining the probabilities of a transition from one note to another. A finer segmentation enables us to analyze the subtle details of the intonation by subdividing the note's pitch contour into a chain of three basic vocal expression segments: steady, transitory and vibrato.

Finally, musical signal visualization methods can be used to represent the 5 dimensions of jingju singing. We list the relevant potential technologies for visualizing each dimension:
- pronunciation: using the formant frequency and amplitude to visualize vowels [2], and spectral centroid, spread, energy to visualize consonants.
- Intonation and loudness: pitch and loudness contours are straightforward to visualize by the PitchYin and Loudness functions in Essentia [7].
- Rhythm: the note, syllable and phoneme onset time positions are the appropriate indicators for disclosing the rhythmic singing problems.
- Tone quality: the harmonics amplitudes are useful for the judgment of the tone quality correctness of vowels.

# 4. Conclusion

In this paper, we first introduced the mouth/heart jingju singing training method and revealed its deficiency-the tutors' verbal feedback is sometimes ambiguous. Then, we claim that MIR technologies could make up this deficiency, thus help the training process. By classifying, finer inspecting the correction occurrences, and conducting a simple survey, we identified the potential MIR technologies for jingju singing training.

This research is non-systematic due to the limited amount of correction occurrence studied and the small number of subjects participated in the survey. A possible future work could be expanding the correction occurrence sample size and involving more survey subjects and re-identifying the technologies.

# Acknowledgement

This by the European Research Council under the European Union's Seventh Framework Program, as part of the CompMusic project (ERC grant agreement 267583).